\documentclass[12pt]{article}
\usepackage{amsfonts}
\usepackage{amssymb,letterspace}
\usepackage{graphics,psboxit,amsmath}

\def\hybrid{\topmargin -20pt    \oddsidemargin 0pt
        \headheight 0pt \headsep 0pt
        \textwidth 6.35in       
        \textheight 9.25in       
        \marginparwidth .875in
        \parskip 5pt plus 1pt   \jot = 1.5ex}

\hybrid

\def\baselinestretch{1.2}

\catcode`\@=11

\def\marginnote#1{}
\newcount\hour
\newcount\minute
\newtoks\amorpm
\hour=\time\divide\hour by60
\minute=\time{\multiply\hour by60 \global\advance\minute by-\hour}
\edef\standardtime{{\ifnum\hour<12 \global\amorpm={am}%
        \else\global\amorpm={pm}\advance\hour by-12 \fi
        \ifnum\hour=0 \hour=12 \fi
        \number\hour:\ifnum\minute<10 0\fi\number\minute\the\amorpm}}
\edef\militarytime{\number\hour:\ifnum\minute<10 0\fi\number\minute}

\def\draftlabel#1{{\@bsphack\if@filesw {\let\thepage\relax
   \xdef\@gtempa{\write\@auxout{\string
      \newlabel{#1}{{\@currentlabel}{\thepage}}}}}\@gtempa
   \if@nobreak \ifvmode\nobreak\fi\fi\fi\@esphack}
        \gdef\@eqnlabel{#1}}
\def\@eqnlabel{}
\def\@vacuum{}
\def\draftmarginnote#1{\marginpar{\raggedright\scriptsize\tt#1}}

\def\draft{\oddsidemargin -.5truein
        \def\@oddfoot{\sl preliminary draft \hfil
        \rm\thepage\hfil\sl\today\quad\militarytime}
        \let\@evenfoot\@oddfoot \overfullrule 3pt
        \let\label=\draftlabel
        \let\marginnote=\draftmarginnote
   \def\@eqnnum{(\theequation)\rlap{\kern\marginparsep\tt\@eqnlabel}%
\global\let\@eqnlabel\@vacuum}  }

\def\preprint{\twocolumn\sloppy\flushbottom\parindent 2em
        \leftmargini 2em\leftmarginv .5em\leftmarginvi .5em
        \oddsidemargin -.5in    \evensidemargin -.5in
        \columnsep .4in \footheight 0pt
        \textwidth 10.in        \topmargin  -.4in
        \headheight 12pt \topskip .4in
        \textheight 6.9in \footskip 0pt
        \def\@oddhead{\thepage\hfil\addtocounter{page}{1}\thepage}
        \let\@evenhead\@oddhead \def\@oddfoot{} \def\@evenfoot{} }

\def\numberbysection{\@addtoreset{equation}{section}
        \def\theequation{\thesection.\arabic{equation}}}

\def\underline#1{\relax\ifmmode\@@underline#1\else
        $\@@underline{\hbox{#1}}$\relax\fi}

\def\titlepage{\@restonecolfalse\if@twocolumn\@restonecoltrue\onecolumn
     \else \newpage \fi \thispagestyle{empty}\c@page\z@
        \def\thefootnote{\fnsymbol{footnote}} }

\def\endtitlepage{\if@restonecol\twocolumn \else \newpage \fi
        \def\thefootnote{\arabic{footnote}}
        \setcounter{footnote}{0}}  

\catcode`@=12
\relax

\def\figcap{\section*{Figure Captions\markboth
        {FIGURECAPTIONS}{FIGURECAPTIONS}}\list
        {Figure \arabic{enumi}:\hfill}{\settowidth\labelwidth{Figure
999:}
        \leftmargin\labelwidth
        \advance\leftmargin\labelsep\usecounter{enumi}}}
 \relax
\def\tablecap{\section*{Table Captions\markboth
        {TABLECAPTIONS}{TABLECAPTIONS}}\list
        {Table \arabic{enumi}:\hfill}{\settowidth\labelwidth{Table
999:}
        \leftmargin\labelwidth
        \advance\leftmargin\labelsep\usecounter{enumi}}}
 \relax
\def\reflist{\section*{References\markboth
        {REFLIST}{REFLIST}}\list
        {[\arabic{enumi}]\hfill}{\settowidth\labelwidth{[999]}
        \leftmargin\labelwidth
        \advance\leftmargin\labelsep\usecounter{enumi}}}
 \relax
\makeatletter
\newcounter{pubctr}
\def\publist{\@ifnextchar[{\@publist}{\@@publist}}
\def\@publist[#1]{\list
        {[\arabic{pubctr}]\hfill}{\settowidth\labelwidth{[999]}
        \leftmargin\labelwidth
        \advance\leftmargin\labelsep
        \@nmbrlisttrue\def\@listctr{pubctr}
        \setcounter{pubctr}{#1}\addtocounter{pubctr}{-1}}}
\def\@@publist{\list
        {[\arabic{pubctr}]\hfill}{\settowidth\labelwidth{[999]}
        \leftmargin\labelwidth
        \advance\leftmargin\labelsep
        \@nmbrlisttrue\def\@listctr{pubctr}}}
 \relax
\makeatother
\newskip\humongous \humongous=0pt plus 1000pt minus 1000pt

\newif\ifdtup

\relax

\def\be{\begin{equation}}
\def\ee{\end{equation}}
\def\ba{\begin{eqnarray}}
\def\ea{\end{eqnarray}}

\def\a{\alpha}

\def\G{\Gamma}

\def\th{\theta}

\def\l{\lambda}

\def\s{\sigma}

\def\bs{\bigskip}
\def\no{\noindent}

\def\IR{\relax{\rm I\kern-.18em R}}
\def\II{\relax{\rm 1\kern-.35em1}}

\renewcommand{\theequation}{\thesection.\arabic{equation}}
\csname @addtoreset\endcsname{equation}{section}

\def\IR{\relax{\rm I\kern-.18em R}}
\def\inv{^{\raise.15ex\hbox{${\scriptscriptstyle -}$}\kern-.05em 1}}

\begin{document}

\begin{titlepage}
\begin{center}

\hfill NEIP-04-06\\
\vskip -.1 cm
\hfill IFT-UAM/CSIC-04-56\\
\vskip -.1 cm
\hfill hep--th/0410022\\

\vskip .5in

{\LARGE Spin chain sigma models with fermions}
\vskip 0.4in

{\bf Rafael Hern\'andez$^{1 \, \dag}$}\phantom{x} and\phantom{x}
 {\bf Esperanza L\'opez}$^2$ 
\vskip 0.1in

${}^1\!$
Institut de Physique, Universit\'e de Neuch\^atel\\
Breguet 1, CH-2000 Neuch\^atel, Switzerland\\
{\footnotesize{\tt rafael.hernandez@unine.ch}}

\vskip .2in

${}^2\!$
Departamento de F\'{\i}sica Te\'orica C-XI
and Instituto de F\'{\i}sica Te\'orica  C-XVI\\
Universidad Aut\'onoma de Madrid,
Cantoblanco, 28049 Madrid, Spain\\
{\footnotesize{\tt esperanza.lopez@uam.es}}

\end{center}

\vskip .4in

\centerline{\bf Abstract}
\vskip .1in
\no
The complete one-loop planar dilatation operator of ${\cal N}=4$ supersymmetric 
Yang-Mills is isomorphic to the hamiltonian of an integrable $PSU(2,2|4)$ quantum 
spin chain. We construct the non-linear sigma models describing the continuum 
limit of the $SU(1|3)$ and $SU(2|3)$ sectors of the complete ${\cal N}=4$ chain. 
We explicitely identify the spin chain sigma model with the one for a superstring 
moving in $AdS_5 \times S^5$ with large angular momentum along the five-sphere.

\noindent

\vskip 2in
\noindent

\footnotesize{$\dag$ Address after October 1$^{\hbox{st}}$: Theory Division, CERN, 
Geneva 23, CH-1211, Switzerland}

\end{titlepage}
\vfill
\eject

\def\baselinestretch{1.2}


\baselineskip 20pt


\section{Introduction}
  
The AdS/CFT correspondence is a fascinating proposal relating the weak coupling 
regime of a gauge theory to a strong coupling regime in string theory, and vice versa. 
A precise formulation of the correspondence would require complete access to the strong 
coupling regime of each theory and, in particular, a detailed understanding of 
the quantization of the string action in an $AdS_5 \times S^5$ background,
which is indeed an 
involved problem. There is however a maximally supersymmetric plane-wave 
limit of the $AdS_5 \times S^5$ background for the IIB string \cite{BFHP}, that can 
be quantized in the light-cone gauge \cite{Metsaev}. On the gauge theory side this limit 
corresponds to focusing on local operators with a large R-symmetry charge $J$, 
of the form $\hbox {Tr} (Z^J \ldots)$, where $Z$ is one of the ${\cal N}=4$ complex 
scalars \cite{BMN}. On the gravity side these operators are described in terms of small 
closed strings with their center of mass moving with large angular momentum $J$ along 
a circle in $S^5$ \cite{GKP}. 

More general string configurations with several large angular momenta along 
$S^5$ were afterwards proposed to correspond to operators composed of the 
three ${\cal N}\!=\!4$ complex scalars \cite{FT}. The energy of these semiclassical 
strings turned out to admit an analytic expansion on the effective parameter 
$\lambda/J^2$, with $\lambda$ the 't Hooft coupling of the gauge theory, and 
thus suggested a direct comparison with the anomalous dimensions of large Yang-Mills operators. 
A serious difficulty for calculating the anomalous dimensions of operators with a 
large number of constituent fields, once quantum corrections are introduced, is operator 
mixing. The way out of this obstacle came from the deep observation that in the planar 
limit the one-loop dilatation operator of ${\cal N}=4$ supersymmetric 
Yang-Mills \cite{dilatation,Beisert} is isomorphic to the hamiltonian of 
an integrable quantum spin chain \cite{MZ,psu}. This identification allows 
to use the powerful Bethe ansatz technique in solving the spectrum of 
anomalous dimensions. Moreover, in the thermodynamic limit of very long spin chains the
algebraic Bethe equations become integral equations, and therefore also
the anomalous dimension of operators with very large 
spin turned reachable. Extending integrability to higher loops 
remains however an open problem, although strong arguments in favor exist
for some sectors of ${\cal N}=4$ \cite{dilatation}, \cite{su23}-\cite{BDS}. 
Using these techniques, the comparison of anomalous dimensions of gauge 
operators and energies of dual semiclassical strings has shown a perfect 
agreement up to two-loops \cite{spinning}-\cite{Ryang}. It should  
however be stressed a disagreement starting at three-loops \cite{spectroscopy,SS,BDS}, which
is currently one of the most intriguing questions on the AdS/CFT correspondence.

A promising path towards a more complete understanding of 
the correspondence between operators with large quantum numbers and 
semiclassical 
string solutions was opened in \cite{Martin}, where the action 
describing the continuum limit of the spin chain in the coherent state basis 
was directly compared with the dual string non-linear sigma model. As a first 
step, the continuum limit of the $SU(2)$ Heisenberg spin chain, associated to the
two-spin holomorphic scalar sector, was shown to reproduce the action of strings 
moving with large angular momentum along an $S^3$ section of $S^5$ \cite{Martin}. 
Equivalence of the spin chain and string non-linear sigma models was then proved 
at two-loop order \cite{KRT}. The identification of both non-linear 
sigma models at the level of the action also implies the equivalence of 
fluctuations around the solutions, and therefore the matching 
holds beyond rigid strings. This approach was subsequently
applied to more general sectors. The non-linear 
sigma model of the integrable $SU(3)$ spin chain, describing the
three-spin holomorphic scalar sector, was found to correspond to 
a string moving with large angular momentum along $S^5$
\cite{su3,Stefanski}, while the non-compact $SL(2)$ 
chain, corresponding to semiclassical strings spinning in both 
$AdS_5$ and $S^5$, was considered in \cite{Stefanski,sl2}.
The continuum limit of the more involved non-holomorphic scalar sector
was shown to reproduce the dual string action on a phase space 
formulation \cite{KT}. 
  
It is important to extend the above studies to the continuum limit 
of gauge theory sectors associated to operators containing also
fermions.
In this way we should reproduce the dual superstring action and not only its
bosonic part, as it was the case in the previous examples. 
A particularly interesting sector of the complete ${\cal N}=4$ chain
is that with spins in the fundamental representation of $SU(2|3)$
\cite{su23}. It describes operators composed of the three complex scalars and 
two of the sixteen gaugino components. This sector is closed at all orders 
in perturbation theory and already presents some of the main characteristics of the higher 
loop dynamics of the complete ${\cal N}=4$ chain, which are absent in the simpler 
$SU(2)$ sector. Namely, the dilatation operator is part of the symmetry algebra and the 
number of sites of the chain is allowed to fluctuate \cite{su23}. In this paper we will 
derive the continuum limit of the $SU(2|3)$ chain.  
We will however restrict our analysis to contributions 
at leading order in the 't~Hooft coupling. This will be enough to exhibit an 
interesting phenomenon present on the string theory side: the coupling of fermions 
to the RR five-form. The plan of the paper is the following. In section 2 we will 
introduce the problem. In section 3 we will construct the non-linear sigma model
for the $SU(1|3)$ spin chain. In section 4 we will recover the spin chain 
sigma model from a large angular momentum limit of the action for a 
superstring rotating in an $S^3$ section of $S^5$ (see also the related developments in \cite{Mikhailov}). 
Section 5 contains the continuum limit of the $SU(2|3)$ chain. We conclude with some directions of 
research in section 6.


\section{${\cal N}=4$ operators with fermions}

The calculation of the one-loop anomalous dimensions of ${\cal N}=4$ supersymmetric 
Yang-Mills single trace operators, in the large N limit, can be very efficiently 
mapped to the problem of finding the spectrum of an integrable spin chain 
\cite{MZ}. The vector space living at each site of the chain depends on the 
sector of Yang-Mills operators that we want to consider. For instance, operators 
composed of the three ${\cal N}=4$ complex scalars map to a ferromagnetic Heisenberg 
$SU(3)$ spin chain, where at each site sits the fundamental
representation. An important characteristic of this sector is that the anomalous dimensions
of long operators only depend on the 't Hooft coupling through
the combination $\lambda \over L^2$, where $L$ is the number of sites of 
the chain or, equivalently, the number of fields that compose the
operator under study. The failure of this property would invalidate the 
derivation of the semiclassical coherent state action describing the
continuum limit of the spin chain as performed in \cite{Martin,KT}. 

We will be interested in the extension of the $SU(3)$ sector to include 
operators with fermions, but preserving the previous important 
property. In the scalar sector, this property does not hold generically
for operators containing both complex scalar fields and their
conjugates, or equivalently, fields with both positive and negative
charges under the three Cartan generators $J_i$ of the $SU(4)$ R-symmetry
group \cite{MZ,KT}. Guided by this fact, we will enlarge the holomorphic
$SU(3)$ scalar sector by allowing for operators with insertions of those 
components of the gaugino field that, as is the case
for the complex scalars, have positive $J_i$ charges. This will reduce 
the sixteen components of the gaugino to two complex combinations
$\lambda_\alpha$, with $\alpha=1,2$. They form a Weyl spinor that transforms as 
$\bf \big(\frac {1}{2} ,0\big)$ under the Lorentz group, and is invariant under 
the $SU(3)$ subgroup of $SU(4)$ \cite{su23}. Thus this sector defines an $SU(2|3)$ spin chain 
based on the fundamental representation ${\bf 3|2}$, which accommodates the three 
scalar fields $Z_i$ and the two fermions $\lambda_\alpha$. The one-loop hamiltonian 
is \cite{su23}
\be
H = \frac {\lambda}{8\pi^2} \sum_{l=1}^L \big( 1-SP_{l,l+1} \big) \; ,
\label{H}
\ee
where $SP_{l,l+1}$ denotes the super-permutation operator between two 
neighboring sites, $l$ and $l+1$.

Our aim in this note will be to derive the non-linear sigma model associated
to the $SU(2|3)$ spin chain, and to compare it to the action for strings on $AdS_5
\times S^5$, including the fermionic sector. This action is very 
involved, already at the classical level, due to the coupling of the 
fermions to the background RR five-form. However, it has been worked 
out explicitely for the fermionic quadratic terms in \cite{MT}. 
The quadratic approximation is valid when the fermionic excitations 
represent a small perturbation over a given bosonic background. 
Such configurations will map on the gauge theory side to operators with
a large number of bosonic fields, and only a few fermionic insertions.
Since the Lorentz group and the R-symmetry group commute, we expect
that long $SU(2|3)$ operators with just a few fermionic insertions will be 
degenerated in the Lorentz index $\alpha$. This implies that, as far
as we will be interested in the comparison with the dual string action
only at the fermionic quadratic level, it will be enough to restrict ourselves
to the $SU(1|3)$ subsector of the $SU(2|3)$ chain. Notice that truncation 
to this subsector is consistent with the one-loop hamiltonian \eqref{H}.

As a warming up step towards $SU(1|3)$, we will first consider the simpler case of 
$PSU(1|1)\subset SU(2|3)$. This subsector is associated to operators of the schematic 
form ${\rm Tr}(Z_1^J \lambda_{1}^{J'})$, with $J+J'\!=\!L$. At each site of this
spin chain sits a two-dimensional vector space with one bosonic and one fermionic generator, 
$|b\rangle$ and $|f\rangle$, respectively. In order 
to construct a discrete sigma model for the spin chain we will introduce a set of coherent 
states at each site of the chain by applying an arbitrary $PSU(1|1)$ rotation to $|b\rangle$,
\ba
&& |n\rangle = e^{i \xi S_y} \, e^{i \chi S_x} \, |b\rangle=
\big(1-{1 \over 2}\, \zeta^\ast \zeta\big) |b\rangle + \zeta \,|f\rangle 
\; , \nonumber \\ 
&&\langle n | = \langle b | \, e^{i S_x \chi} \, e^{i S_y \xi} =
\langle b | \big(1-{1 \over 2}\, \zeta^\ast \zeta\big) + \langle f |\,
\zeta^\ast \; , \label{cs}
\ea
where $S_x$ and $S_y$ are the two odd generators of $PSU(1|1)$,
$\xi$ and $\chi$ are two Grassmann variables, and we have introduced the complex 
combination $\zeta=-\xi+i\chi$. The coherent states \eqref{cs}
verify
$\langle n | n \rangle=1$. 
They form an overcomplete basis, with the resolution of the identity 
\be
\int d\zeta d\zeta^\ast \, |n\rangle \langle n| = \II \ .
\label{id}
\ee
  
A path integral description of the partition function, as an integral over the 
overcomplete set of coherent states, shows the equivalence between the spin
chain and the following discrete sigma model (see for instance \cite{FA} for 
a complete derivation and details on the simpler case of the $SU(2)$ spin $s$ chain)
\be
S = - \int dt \, \Big[\, i \langle {\bf n} |{d \over dt}|{\bf n} \rangle + 
\langle {\bf n} |H|{\bf n} \rangle \Big] \; ,
\label{ds}
\ee
where we have introduced the product of coherent states along the chain 
$|{\bf n} \rangle=|n_1 \cdots n_L\rangle$. 

In order to derive the effective action describing the continuum limit
of the spin chain, it is generically necessary
to take into account quantum effects 
of short wavelength configurations \cite{KRT}. However, when we are only 
interested 
in capturing the physics at first order in ${\lambda \over L^2}$, it is 
enough to substitute in \eqref{ds} finite
differences between variables at neighboring sites
by derivatives \cite{FA}. After an straightforward calculation we obtain  
\be
S = - {L \over 2 \pi}\int d\s dt \, \Big[ \, i \zeta^\ast \partial_t \zeta + 
\frac {\l}{2L^2}
\partial_\sigma \zeta^\ast \partial_\sigma \zeta \, \Big] \; ,
\label{psu11}
\ee
which is the action describing a non-relativistic fermion, with mass
$m={L^2 \over \lambda}$. This action reproduces, at first order
in $\lambda/L^2$, the spectrum of fermionic fluctuations in the 
$pp$-wave limit \cite{BMN}.


\section{The $SU(1|3)$ spin chain}

The $SU(1|3)$ sector of ${\cal N} \! = \! 4$ Yang-Mills consists of
operators composed of the 
three complex scalar fields and one component of the Weyl fermion. 
The derivation of the continuum limit of the associated
$SU(1|3)$ spin chain follows the same steps as above.
We will introduce a set of spin coherent states
at each site of the chain by applying an arbitrary
rotation to a bosonic eigenstate of the Cartan generators.
Since such state is left invariant by an $SU(1|2)$ subgroup and
just multiplied by a phase via an additional $U(1)$, the set of 
coherent states will be isomorphic to the quotient $SU(1|3)/SU(1|2)\times U(1)$. 
Hence to describe them we need four real variables and two Grassmann ones. 
An easy way to construct the $SU(1|3)$ coherent states is to consider first 
those for a $PSU(1|1)$ subsector and then apply to them an arbitrary $SU(3)$ 
rotation. The resulting states will have the same form \eqref{cs} as
before, but with 
$|b\rangle$ denoting now a $SU(3)$ coherent state (we will use the same 
notation and conventions as in \cite{su3}), 
\be
|b\rangle =  
\cos \theta \, \cos \psi \, e^{i \varphi} \, |1 \rangle + 
\cos \theta \, \sin \psi \, e^{-i \varphi} \, |2 \rangle + 
\sin \theta \, e^{i \phi} \, |3 \rangle\ \ ,
\label{su3}
\ee
where $\theta, \psi \in [0,\pi/2]$ and 
$\phi+\varphi, \phi-\varphi \in [0,2\pi]$. 
The coherent states satisfy again $\langle n|n
\rangle=1$ and expand the identity in an expression analogous to \eqref{id}.

As before, the spin chain system is equivalent to a discrete sigma model
with action \eqref{ds}. 
The continuum limit of the chain is obtained by evaluating
\eqref{ds} over configurations that vary smoothly along the chain.
The calculation is lengthy but simple, and we obtain
\be
S=S_B+S_F \; .
\label{St}
\ee
The bosonic part of the action is the $SU(3)$ spin chain sigma model
derived in \cite{su3,Stefanski}
\be
S_B=\frac {L}{ 2 \pi} \int d\s \, dt \, \big[\, C_0-\frac {\l}{2L^2}\,
e^2\, ] \; , 
\label{Sb}
\ee
where the Wess-Zumino term, $C_0$, is the time component of the 1-form
\be
C= -i \langle b|d|b \rangle = \sin^2 \theta \; d \phi + \cos^2 \theta \, 
\cos (2 \psi) \; d\varphi \; ,
\label{conn}
\ee
and we have defined
\be
e^2 = \theta'^2 
+ \cos^2 \theta \, \left( \psi'^2 + \sin^2 (2 \psi) \, \varphi'^2  \right)
+ \sin^2 \theta \, \cos^2 \theta \, \left( \phi'- \cos (2 \psi) \, \varphi' 
\right)^2 \; , 
\label{defe}
\ee
which represents energy density of the $SU(3)$ sigma model,
$H={\lambda \over 4 \pi L} \int d\sigma \, e^2$.
The fermionic piece of the action is given by
\be
S_F = - \frac {L}{2\pi} \int d\s dt \, \Big[ \, i \zeta^* D_t \zeta + 
\frac {\l}{2L^2} \Big( D_{\s} \zeta^* D_{\s} \zeta 
- e^2 \zeta^\ast \zeta \Big) \, \Big] \ ,
\label{e}
\ee
with $D\zeta = d\zeta -i C\zeta$. One more condition should be added 
to \eqref{St} in order to 
represent Yang-Mills operators. The trace required to define gauge 
invariant operators implies that we should only considered
translationally invariant field configurations, {\it i.e.} $P_\sigma=
2\pi {\mathbb Z}$.

Definitions of the coherent states that differ by a phase 
should be considered equivalent. Therefore in order to parameterize
the set of coherent states, we have to make a choice of global phase. 
Under a change in the definition $|n\rangle \rightarrow e^{i\alpha} |n\rangle$, 
we have $\zeta \rightarrow e^{i \alpha} \zeta$ and $C \rightarrow C+d\alpha$.
Consistently, this transformation leaves invariant $S_F$ and only changes 
the bosonic term $S_B$ by an irrelevant total derivative. However, it is 
important to notice that a phase choice can not be fixed globally over the set
of $SU(3)$ coherent states. In particular, the one implicit in \eqref{su3} becomes 
singular at $\theta={\pi \over 2}$ and $\psi=0, {\pi \over 2}$. The appearance of 
the covariant derivatives in \eqref{e} is a consequence of the non-triviality of the 
line bundle $\{e^{i \alpha} |b\rangle\}$, contrary to the simpler $PSU(1|1)$ case.


\section{The superstring action}

In this section we will describe how the fermionic quadratic terms of the
dual superstring action reproduce, after some large angular 
momentum limit, the spin chain results obtained in the previous section.
The fermionic part of the type IIB Green-Schwarz superstring 
action in $AdS_5 \times S^5$ expanded to quadratic order near a particular bosonic 
string solution (with a flat induced metric) was described in 
\cite{MT}. Choosing the $\kappa$-symmetry so that both Majorana-Weyl
spinors in ten dimensions are equal, the quadratic fermionic lagrangian is
\be
L =  - 2i \, \bar{\vartheta} \Big( \rho^a D_a + {i \over 2} \epsilon^{ab} \rho_a 
\Gamma_\ast \rho_b \Big) \vartheta \;\; , \hspace{.5cm} D_a=\partial_a + {1
\over 4} \omega_a^{AB} \Gamma_{AB} \; ,
\label{Sfermions}
\ee
where $\rho_a$ and $\omega_a^{AB}$ (with $a=0,1$) are projections of 
the $AdS_5 \times S^5$ gamma matrices and spin connection,
\be
\rho_a= \partial_a X^M E^A_M \, \Gamma_A \;\; , \hspace{.5cm} 
\omega_a^{AB}= \partial_a X^M \omega_M^{AB} \; ,
\ee
and $X^M$ denotes the coordinates on $AdS_5$ (when $M=(0,6,7,8,9)$) and
$S^5$ (for $M=(1,2,3,4,5)$), $E_M^A$ is the ten-vein, and $\Gamma_A$ are 
the flat space ten-dimensional gamma matrices. The second term in (\ref{Sfermions}) is a 
mass term, and has its origin in the coupling to the RR 5-form \cite{MT}; we have defined 
$\Gamma_\ast= i \Gamma_{06789}$, with $\Gamma_\ast^2=\II$.
  
The action for relativistic bosonic strings rotating in $S^5$ has been 
shown to reproduce the $SU(3)$ spin chain sigma model \eqref{Sb} 
\cite{su3,Stefanski}. Hence, in order to reproduce the fermionic piece of 
the $SU(1|3)$ spin chain action, we should consider such string solutions 
as the bosonic background for the quadratic action \eqref{Sfermions}.
However, to simplify the calculation while keeping at a general level, 
we will restrict to strings rotating in an $S^3$ section of $S^5$. 
These bosonic solutions map to large operators of ${\cal N}=4$ supersymmetric 
Yang-Mills composed of two of the three complex scalars, and 
correspond to setting $\theta=0$ in \eqref{Sb}. The metric on
${\mathbb R}_t \times S^3$ is
\be
ds^2= -dt^2 + d \psi^2 + \cos^2 \! \psi \, d \phi_1^2 + \sin^2 \!\psi \,
d \phi_2^2
\, .
\ee
We will make the gauge choice $t=\kappa \tau$, and associate the tangent
space labels $A=0,1,2,3$ with the coordinates $t,\psi,\phi_1,\phi_2$ 
respectively. Then (with $\dot{X}^M=\partial_t X^M$ 
and ${X'}^M=\partial_{\s} X^M$ for the time and space derivatives)
\ba
&& \rho_0 = \kappa \Big[ \G_0 + \dot{\psi} \G_1 + \dot{\phi}_1 \cos \psi 
\G_1 + \dot{\phi}_2 \sin \psi \G_3 \Big] \, , \nonumber \\
&& D_0 = \kappa \Big[ \partial_t + {1\over 2} \dot{\phi}_1 \sin \psi \G_{12}
-{1\over 2} \dot{\phi}_2 \cos \psi \G_{13} \Big] \, , \nonumber \\
&& \rho_1 = {\psi}' \G_1 + \phi_1' \cos \psi 
\G_2 + \phi'_2 \sin \psi \G_3  \, , \\
&& D_1 = \partial_\sigma + {1\over 2} \phi'_1 \sin \psi \G_{12}
-{1\over 2} \phi'_2 \cos \psi \G_{13} \nonumber \, .
\ea

While the spin chain action \eqref{e} is a one-loop result, the string action 
\eqref{Sfermions} captures arbitrary orders of the effective coupling 
constant. For comparing both actions, it is thus
enough to take into account the effect of the rotating string 
background on the fermionic fluctuations at order $\lambda/L^2$.
The energy of the bosonic string solutions that we 
are considering verifies \cite{FT}
\be
E=\sqrt{\lambda} \, \kappa = L \left[ 1 + {\cal O}\Big( {\lambda \over
L^2} \Big) \right] \; ,
\ee
where $L$ denotes the total angular momentum.
Since $\kappa$ and $L/\sqrt{\lambda}$ only differ in subleading
terms, and we are only interested 
in the leading dependence on the coupling constant, we can use 
$\kappa$ instead of $L/\sqrt{\lambda}$ as the expansion parameter of
the string action. 
A systematic expansion can be obtained after we  
distinguish between fast and slow movement of the rotating string
solution \cite{Martin,KT}.
This is facilitated by the change of coordinates
\be
\phi_1=\alpha+\varphi \;\; , \hspace{5mm} \phi_2=\alpha-\varphi \; .
\label{sv}
\ee
The conjugated momentum of the variable $\alpha$ is the total angular 
momentum $L$. The limit of large $L$ corresponds
to strings rotating close to the speed of light 
in the $\a$ direction. This fast movement can be absorbed by shifting
$\alpha \rightarrow \alpha +t$. After this redefinition, the string
propagation verifies $\dot{X}^M = 
{\cal O}\big({1 \over \kappa^2}\big)$, for $M \neq 0$ \cite{Martin}.

The spin chain action for the fermionic fluctuations \eqref{e} describes 
two-dimensional fermions with canonically normalized kinetic terms, coupled to a gauge 
field which carries the information about the bosonic background.
Contrary, in the fermionic string action \eqref{Sfermions} the background 
is felt both on the matrices $\rho_a$ and on the covariant derivatives $D_a$. 
For the comparison of both actions, the first step is to trivialize the matrices 
$\rho_a$ by applying a series of rotations and concentrate the effect of the 
background only on $D_a$ \footnote{The same process was followed in 
\cite{semi} to derive the spectrum of fermionic fluctuations around a
particular circular string solution.}. 
Let us start by writing the matrices $\rho_a$ in terms 
of the variables \eqref{sv}, after the shift $\a \rightarrow \a + t$, 
\ba
&& \rho_0=\kappa \Big[ \G_0+ 
\dot{\psi} \G_1 + (1+\dot{\alpha}) (\cos \psi \G_2 + \sin \psi \G_3)
+ \dot{\varphi} (\cos \psi \G_2 - \sin \psi \G_3) \Big] \; , \nonumber \\
&& \rho_1= \psi' \G_1 + \alpha' (\cos \psi \G_2 + \sin \psi \G_3)
+ \varphi' (\cos \psi \G_2 - \sin \psi \G_3) \; .
\ea
We can now perform the necessary rotations $\rho_a \rightarrow S \rho_a S^{-1}$ 
as an expansion in $1/\kappa^2$, from which we will only retain the leading 
term. Applying two consecutive rotations $S_1=e^{{1 \over 2} p_1 \G_{23}}$ 
and $S_2=e^{{1 \over 2} p_2 \G_{21}}$, with
\be
p_1= \psi - \sin(2\psi)  \dot{\varphi} \; \; , \hspace{5mm} 
p_2= \dot{\psi} \; ,
\ee
we obtain
\be
\rho_0= \kappa \Big[ \G_0 + \Big(1 - {e^2 \over 2 \kappa^2}\Big) \G_2 \Big]
\; \; , \hspace{5mm}  \rho_1= \psi' \G_1 - \sin(2 \psi) \varphi' \, \G_3 \; ,
\ee
where we have used the Virasoro constraints, and $e$ is the quantity
defined in \eqref{defe}. Two further rotations lead to 
$\rho_0=e\G_0$ and $\rho_1=e\G_3$,
\ba 
&& S_3=e^{{1 \over 2} p_3 \G_{02}} \;\; , \hspace{5mm} 
\cosh p_3= {\kappa \over e} \;\; , \hspace{5mm} 
\sinh p_3 = -{\kappa \over e} + {e \over 2 \kappa}  \, , \nonumber \\ 
&& S_4=e^{{1 \over 2} p_4 \G_{31}} \;\; , \hspace{5mm} \cos p_4=- 
{\sin(2 \psi) \varphi'
\over e} \;\; , \hspace{5mm} \sin p_4={\psi' \over e} \, .  
\ea
  
The transformation of the covariant derivatives $D_a$ under the previous 
rotations is tedious but straightforward to derive. The final result for 
the quadratic fermionic lagrangian \eqref{Sfermions} is
\be
L=-2i \kappa^2 \, \bar{\Psi} \left[\G_0 \big(\partial_t +i \Pi 
A_t \big)- {1 \over \kappa}\G_3 \big(\partial_\sigma +i \Pi A_\sigma \big) +
\Big(1-{e^2 \over 2\kappa^2}\Big) \G_0 \G_3 \G_2 \bar{\Pi} \right] \Psi \; ,
\label{Lfermions}
\ee
with $\Pi=-i\G_0\G_1\G_2\G_3$ and $\bar{\Pi}=\G_6\G_7\G_8\G_9$. The
new spinor field is given by 
$\Psi= \sqrt{e \over \kappa}\, S_4 S_3 S_2 S_1 \, \vartheta$.
The factor ${\kappa}^{-{1 \over 2}}$ in the normalization of the spinor 
has been introduced to compensate the effect of the $S_3$ rotation, such that 
$\Psi$ remains finite in the large $\kappa$ limit. We have also introduced 
\be
A_t = C_t+ {1 \over 2}(\dot{p}_4-1)  \;\; , \hspace{5mm} 
A_\sigma=C_\sigma+{1 \over 2} \, p'_4 \; ,
\label{con}
\ee
with $C$ as defined in \eqref{conn}, with $\th=0$. This background field
can be interpreted as a connection for the transformation
$\Psi \rightarrow e^{{i \over 2} \Pi f } \Psi$, which is a symmetry
of the action \eqref{Lfermions};
choosing $f=t-p_4$ we obtain 
\be
A=C \; .
\ee

It was important in deriving \eqref{Lfermions} to separate the movement
of the background $S^3$ string solution into fast and slow components,
since this allowed us to perform a $\kappa$-expansion of the action.
For consistency, in the time evolution of the fermions we should also 
distinguish between fast and slow oscillations. This can be achieved 
in the following way. As in \cite{semi}, we split the ten-dimensional spinor $\Psi$ 
into eigenstates of the projector ${1\over 2} (1+\bar{\Pi})$, which commutes with all 
other operators. Given that the gamma matrices corresponding to two of
the $S^5$ directions do not appear in \eqref{Lfermions}, we can divide 
the eight real components of $\Psi$ with the same $\bar{\Pi}$-eigenvalue 
into two sets of four which do not mix with each other. We can now interpret 
\eqref{Lfermions} as the lagrangian for a spinor with four real 
components, and consider $\G_M$ as four-dimensional gamma matrices, with $M=0,..,3$. 
We will choose the representation 
$\G_0=i\binom{\,0\;1\,}{\,1\;0\,}$ and $\G_j=i
\binom{\, 0 \, - \sigma_j\!}{\sigma_j \;\;0}$, 
with $\sigma_1\!=\!\sigma_z$, $\sigma_2\!=\!\sigma_x$, 
$\sigma_3\!=\!\sigma_y$ the Pauli matrices. The dynamics defined by 
\eqref{Lfermions} is then consistent with taking $\Psi$ to have the structure of a
four-dimensional Majorana spinor, $\Psi=\binom{\xi}{\sigma_y \xi^\ast}$. 
The last term in \eqref{Lfermions} gives mass to $\xi=\binom{\xi_1}{\xi_2}$, 
with $m\!=\!\pm \big(1-{e^2\over 2 \kappa^2}\big)$ for each component. 
The fast oscillations of the fermionic fluctuations are due to
the contribution $\pm 1$ to $m$, and can be subtracted from the
positive frequency components by redefining $\xi \rightarrow
e^{it} \xi$. After this shift, the equations of motion become
\ba
&& D_t \, \xi_1 +{i \over \kappa}D_\sigma \, \xi_2
+{i e^2\over 2 \kappa^2} \,  \xi_1=0 \, , \\ \nonumber
&& D_t \, \xi_2 -{i \over \kappa} D_\sigma \, \xi_1
- {ie^2 \over 2 \kappa^2} \, \xi_2= - 2 i \, \xi_2 \, ,
\ea
with $D=d-iC$. At leading order in $1/\kappa$ they imply
\be
\xi_2={1 \over 2\kappa} D_\sigma \, \xi_1 \; .
\ee
Substituting in \eqref{Lfermions}, the action becomes 
\be
S = - \frac {\sqrt{\lambda} \kappa}{2 \pi} \int d\s dt \, \left[\, i\xi_1^\ast D_t  \xi_1 +
{1 \over 2\kappa^2} \Big( D_\sigma \xi_1^\ast D_\sigma \xi_1 - 
e^2 \xi_1^\ast \xi_1 \Big) \right] \, .
\label{Sstring}
\ee
This action precisely coincides with the spin chain action (\ref{e}) in 
the $\th=0$ sector, after we use the leading order relation 
$\kappa=L/\sqrt{\lambda}$. Notice that the term proportional to $e^2$
has its origin, from the string theory point of view, in the coupling of
the fermions to the background RR 5-form. 

From the ten-dimensional spinor $\Psi$ we actually obtain four
fermions with the same action \eqref{Sstring}. 
These fluctuations around the $\th=0$ sector, or equivalently,
around a bosonic string rotating on a $S^3$ section of $S^5$, map
to long Yang-Mills operators composed of two complex scalars with
few fermionic insertions. From the sixteen gaugino components
there are four of them positively charged under the two Cartan 
generators of the $SU(4)$ R-symmetry group
that act non-trivially on two complex scalars. Following the
arguments of section 2, we expect then a four-fold degeneracy
also in the gauge theory side.

The comparison between the
superstring action, including the fermionic part, and a dynamical
system describing long Yang-Mills 
operators was also addressed in \cite{Mikhailov} following a different 
approach.


\section{The $SU(2|3)$ spin chain}

In this section we will derive the continuum limit of the 
complete $SU(2|3)$ spin chain. As we have explained in section 2,
the $SU(2|3)$ sector of ${\cal N} \! = \! 4$ is associated to operators composed of
the three complex scalars and two complex fermions. We will choose the following 
parameterization for the coherent states in this sector,
\be
|n\rangle = \big( 1 - \frac {1}{2} \zeta_1^* \zeta_1 - \frac {1}{2} 
\zeta_2^* \zeta_2 
+ \frac {3}{4} \zeta_1^* \zeta_1 \zeta_2^* \zeta_2 \big) 
\big( |b\rangle + \zeta_1 |f_1\rangle + \zeta_2 |f_2\rangle \big) \ ,
\label{su23}
\ee
with $|b\rangle$ again the $SU(3)$ coherent state (\ref{su3}), 
$|f_1\rangle$ and $|f_2\rangle$ fermionic 
states belonging to the fundamental representation of $SU(2|3)$, and 
$\zeta_1$ and $\zeta_2$ two complex Grassmann variables. 
The term factored out in \eqref{su23} insures $\langle n |n\rangle=1$.

The action describing the continuum limit of the chain is obtained
following the steps by now familiar. The bosonic part of this action
gives again the $SU(3)$ non-linear sigma model \eqref{Sb}. For the
fermionic part we obtain
\ba
S_F & = & - \frac {L}{2\pi} \int d\s dt \, \Big\{ \, i \zeta_1^* D_t \zeta_1 + 
i \zeta_2^* D_t \zeta_2 - i z^* D_t z  +  \\ \label{S23}
    & + & \!\! \frac {\l}{2L^2} \Big[ D_{\s} \zeta_1^* D_{\s} \zeta_1 
+D_{\s} \zeta_2^* D_{\s} \zeta_2 - D_{\s} z^* D_{\s} z - e^2 
\big(\zeta_1^\ast \zeta_1 +\zeta_2^* \zeta_2 -
2 z^* z\big) \Big]\, \Big\} \ , \nonumber
\ea
where we have defined 
\be
z=\zeta_1 \zeta_2 \; ,
\ee
and the covariant derivatives are $D\zeta_j = d\zeta_j -i C\zeta_j$ 
(with $j=1,2$) and $D z = dz -2i C z$. 
The action \eqref{S23} describes two non-relativistic fermions 
with quartic interactions. At the quadratic level this action coincides with \eqref{e}
for each of the fields $\zeta_j$. This confirms that it is enough to 
study the $SU(1|3)$ subsector when we are only interested in the
quadratic fermionic terms. It would be very interesting to compare the
quartic terms obtained here with the dual superstring action.


\section{Conclusions}

In this paper we have derived the non-linear sigma model arising from 
the continuum limit of the spin chains associated to the $SU(1|3)$ and 
$SU(2|3)$ sectors of ${\cal N}=4$ Yang-Mills. We have then explicitely 
recovered the $SU(1|3)$ spin chain sigma model from the non-linear sigma model 
for a superstring rotating in an $S^3$ section of $S^5$, showing thus
that the identification of field and string theory actions also holds
when fermions are included.
  
We have restricted our analysis to leading order in the 't Hooft 
coupling. In \cite{su23} the dilatation operator in the $SU(2|3)$ sector 
has been determined, and shown to be integrable, up to three-loops.
It would be interesting then to extend our derivation to higher
loops following \cite{KRT}, where the comparison between the spin chain 
and the string actions was performed to two-loops in the $SU(2)$ sector.
At higher order in the 't Hooft coupling, the $SU(2|3)$ sector presents 
a feature characteristic of the complete ${\cal N}=4$ theory: the number
of sites is allowed to fluctuate, due to the mixing of the three scalar
fields into two fermions. This mixing has been argued to be absent in 
the thermodynamic limit of the chain \cite{beyondsu2}, making the
extension of our analysis to two-loops feasible. To address the 
important phenomenon of length fluctuation, it would be necessary to consider 
$1/L$ corrections. It should be noted however that the comparison of the $1/L$ 
corrections between the gauge and gravity sides is an open problem, even at 
leading order \cite{zL,tL}. A relevant result for understanding what is the reflect on the
string side of the fluctuations in length of the spin chain could come from \cite{mikha}, 
where a certain conserved charge of generic fast moving strings has 
been conjectured to be associated with the spin chain length.
  
Another interesting direction to follow is analyzing the continuum 
limit of spin chains with less symmetries. Sigma models describing
strings on orbifolds have recently been shown to match with the 
continuum limit of spin chains with twisted boundary conditions \cite{Ideguchi}.
Further cases to study are the spin chain for 
${\cal N}=2$ Yang-Mills \cite{DiVecchia}, or chains connected to deformations of ${\cal
N}=4$ Yang-Mills \cite{Roiban,BC}. This approach could provide 
a constructive way of obtaining the string duals of more generic gauge 
theories and, at the same time, understanding the behavior of string
theory in more general backgrounds.


\bs\bs

\centerline {\bf Acknowledgments}

It is a pleasure to thank E. \'Alvarez, C. G\'omez, K. Landsteiner and 
T. Ort{\'\i}n for useful discussions. 
R.H. acknowledges the financial support provided through the European
Community's Human Potential Programme under contract HPRN-CT-2000-00131
``Quantum Structure of Space-time'', the Swiss Office for Education 
and Science and the Swiss National Science Foundation. The work of E.L. 
was supported by a Ram\' on y Cajal contract of the MCYT and 
in part by the Spanish DGI under contract FPA2003-04597.


\end{document}